\documentclass[aps,prb,reprint,superscriptaddress,floatfix]{revtex4-1}

\usepackage{epsfig}
\usepackage{units}

\begin{document}


\title{Damping of parametrically excited magnons in the presence of the longitudinal spin Seebeck effect} 



\author{Thomas Langner}
\email[]{tlangner@rhrk.uni-kl.de}
\affiliation{Fachbereich Physik and Landesforschungszentrum OPTIMAS, Technische Universit\"at Kaiserslautern, 67663 Kaiserslautern, Germany}

\author{Akihiro Kirihara}
\affiliation{IoT Devices Research Laboratories, NEC Corporation, Tsukuba 305-8501, Japan}

\author{Alexander A. Serga}
\affiliation{Fachbereich Physik and Landesforschungszentrum OPTIMAS, Technische Universit\"at Kaiserslautern, 67663 Kaiserslautern, Germany}

\author{Burkard Hillebrands}
\affiliation{Fachbereich Physik and Landesforschungszentrum OPTIMAS, Technische Universit\"at Kaiserslautern, 67663 Kaiserslautern, Germany}

\author{Vitaliy I. Vasyuchka}
\affiliation{Fachbereich Physik and Landesforschungszentrum OPTIMAS, Technische Universit\"at Kaiserslautern, 67663 Kaiserslautern, Germany}

\date{\today}

\begin{abstract}
The impact of the longitudinal spin Seebeck effect (LSSE) on the magnon damping in magnetic-insulator/nonmagnetic-metal bilayers was recently discussed in several reports. However, results of those experiments can be blurred by multimode excitation within the measured linewidth. In order to avoid possible intermodal interference, we investigated the damping of a single magnon group in a platinum covered Yttrium Iron Garnet (YIG) film by measurement of the threshold of its parametric excitation. Both dipolar and exchange spin-wave branches were probed. It turned out that the LSSE-related modification of spin-wave damping in a micrometer-thick YIG film is too weak to be observed in the entire range of experimentally accessible wavevectors. At the same time, the change in the mean temperature of the YIG layer, which can appear by applying a temperature gradient, strongly modifies the damping value.

\end{abstract}


\maketitle 


Spin caloritronics, the research field focused on the interaction between magnetic and thermal effects, has attracted a lot of interest in recent investigation activities \cite{Bauer2012, Boona2014, Tserkovnyak2016}. The possibility to control and manipulate magnetic processes by thermal means offers a high potential for application. Furthermore the finding of ways to reinvest waste heat is one of the major challenges towards green energy techniques \cite{Kirihara, Kirihara2016}.
The spin Seebeck effect is one of the most fascinating phenomena in this research area. Here a spin current is generated by a temperature gradient across the interface between a magnetic material and a nonmagnetic metallic layer \cite{Uchida2010}. Usually the inverse spin Hall effect is used to detect this spin current\cite{SaitohISHE}. This means that it is converted into a charge current due to spin orbit interaction inside the nonmagnetic metal and, thus, an electric voltage can be detected. Since its observation the spin Seebeck effect has become the major tool in spincaloritronic research. Although there has been much effort to reveal the nature of the spin Seebeck effect\cite{Schreier2016, Kehlberger2015} and to develop possible application schemes, there are still open questions. For example, the influence of the spin Seebeck effect on the magnetization dynamics is still under discussion.

Recent research activities were focused to find evidence that the spin Seebeck effect can partially compensate magnetic damping\cite{Jungfleisch2013, Lu2012} and can even enhance the magnetization precession\cite{Rezende1} in case of dipolar spin waves. It has been reported that a generated spin current establishes an additional spin torque to the ferromagnet\cite{Otani2014}. If the temperature of the non-magnetic metal layer is higher than the temperature of the magnetic material a spin angular momentum is transferred into the ferromagnet and reduces the effective damping. In the opposite case this angular momentum is absorbed by the nonmagnetic metal layer and thus the effective magnetic damping is enhanced. In order to gain more insight into this possible influence on the damping of excited magnons we investigate the threshold power of the parametric generation of magnons.
Parametric pumping is a well established, powerful tool to excite and amplify magnons in the dipolar and in the dipolar-exchange areas of a spin-wave spectrum \cite{Schloemann, Gurevich, Zakharov}. Hereby an alternating magnetic field oscillating with twice the magnon frequency is applied parallel to the magnetization of a ferromagnet. A microwave frequency photon converts into two magnons with half the frequency and opposite wavevectors. If the energy transferred to the spin system overcomes the spin-wave losses, parametric instability occurs and the magnon density increases exponentially with time. Thus the parametric magnon generation is a threshold process. An influence of the spin Seebeck effect on the damping will modify this pumping threshold.
Therefore we investigate the impact of a temperature gradient on a magnonic system in a bilayer of a ferrimagnetic film and an attached nonmagnetic metallic film with high spin orbit coupling.
We show that this influence of a temperature gradient is weak and cannot be detected within the experimental errors. We compare these results with homogeneous variation of the temperature. In that case the influence on the damping is pronounced.


\begin{figure}[h]
\includegraphics[width=0.9\linewidth]{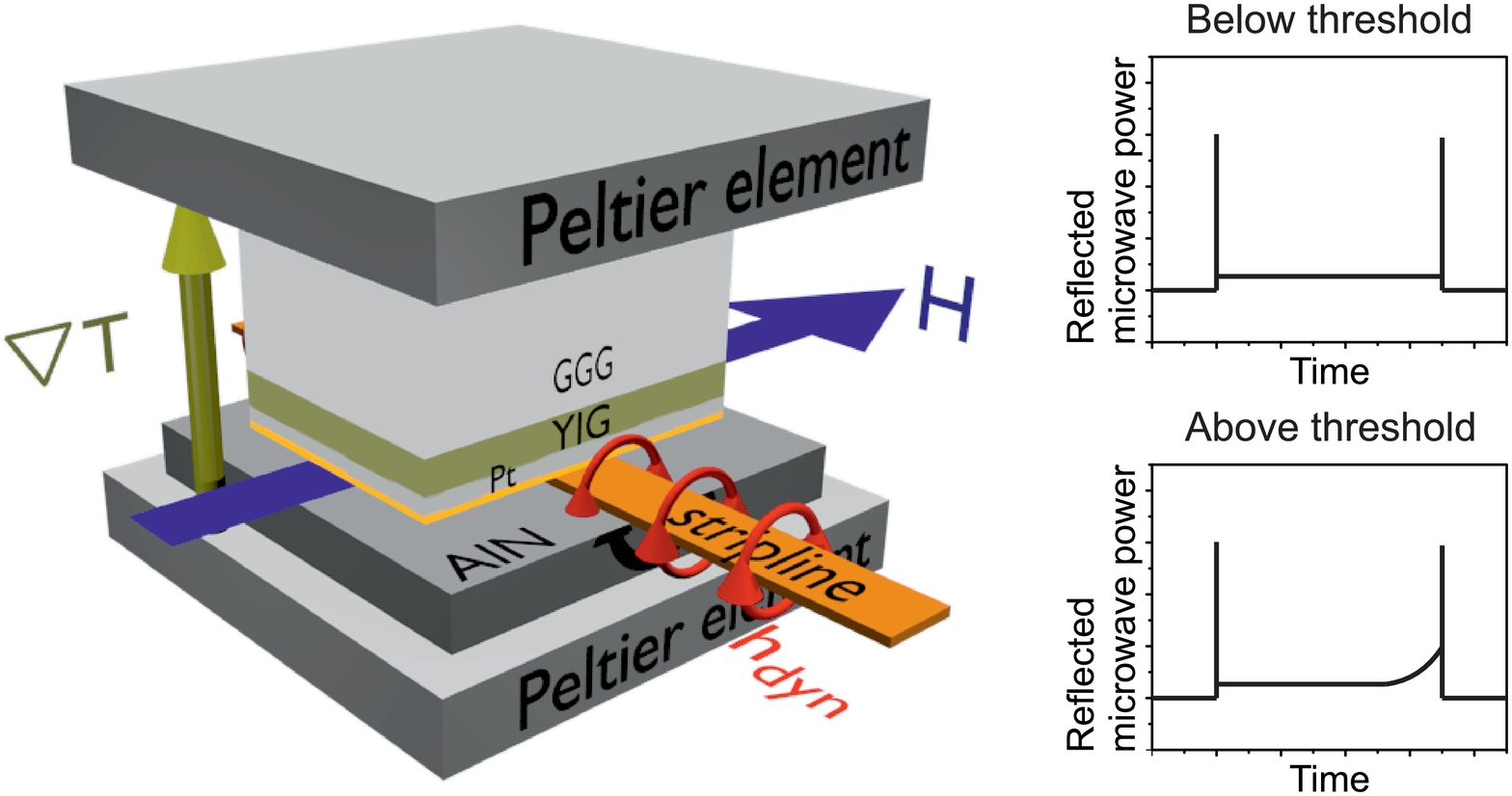}%
\caption{\label{image}Scheme of the experimental setup. The Pt/YIG/GGG sample and the AlN substrate with microstrip line are clamped between two separately controlled Peltier elements that create a thermal gradient $\nabla T$ across the sample. The Pt layer is electrically isolated from the copper microstrip by a thin PMMA coating. The microstrip is part of a microwave resonator. A microwave current applied to the resonator creates a dynamic Oersted field ${\mathbf{h}}_{\mathrm{dyn}}$ with a component parallel to the static magnetization of the sample. The whole setup is mounted in a heat sink. On the right side the measurement scheme is shown. The reflected signal delivers information about the creation of magnons.}%
\end{figure}

In order to perform the mesaurements, the experimental setup shown in Fig.~\ref{image} is used. The investigated sample is a multilayer of a 5\,nm thick Platinum (Pt) film, a 6.7\,${\mu}$m thick Yttrium Iron Garnet (YIG) film and a 500\,${\mu}$m thick Gallium Gadolinium Garnet (GGG) substrate. The YIG layer is grown on the GGG substrate in (111)-orientation by liquid phase epitaxy. The platinum film was sputtered on the YIG surface. With the platinum layer at the bottom, the sample is placed on top of a 50\,$\mu$m wide copper microstrip. A thin coating layer of Polymethyl Methacrylate (PMMA) electrically isolates the Pt layer from this microstrip. The microstrip is structured on a metallized aluminum nitride (AlN) substrate that is used due to its high thermal conductivity at room temperature \cite{slack} of 285\,W\,m$^{-1}$K$^{-1}$. Two separately controlled Peltier elements, one below the AlN substrate and one directly on top of the sample, are used to create a temperature gradient perpendicular to the sample plane, see Fig.~\ref{image}. Using Peltier elements the temperature configurations can be precisely controlled for uniform temperatures and also for a temperature gradient in both directions. Nevertheless the maximal possible temperature difference in both directions is limited to about 20\,$^{\circ}$C in the experimental setup at around room temperature due to heat dissipation. The back sides of the Peltier elements are connected to heat sinks that are clamped between the water-cooled poles of an electromagnet ensuring effective heat transfer in the system. The externally applied magnetic field {\bf H} is oriented in plane of the sample and perpendicular to the long axis of the copper stripline.\\

A stub tuner connected to the microstrip line creates a tunable microwave resonator. 10\,${\mu}$s long microwave pulses with a carrier frequency of 14\,GHz and a repetition time of 10\,ms applied to this resonator create an alternating Oersted field ${\mathbf{h}}_{\mathrm {dyn}}$ around the microstrip. The small width of the microstrip leads to a high microwave current density and, thus, a strongly localized high magnetic field density. The reflected microwave signal from this resonator is rectified by a semiconductor diode and the envelope is shown on an oscilloscope. The microwave power applied to the resonator is controlled to exactly that level where the reflected pulse starts showing a kink at the end of the pulse profile. This kink appears as a consequence of a change in the quality factor of the tuned resonator due to the excitation of magnons, and thus gives evidence for the appearance of the parametric instability \cite{Sandweg2011, Neumann2009}. In this case the applied microwave power level is considered as the threshold power. Since the amount of generated magnons is still low at the threshold power level and the pumping pulse is switched off close to this point, we can neglect the additional heating of the sample by magnon-phonon transfer.

\begin{figure}[t]
\includegraphics[width=0.90\linewidth]{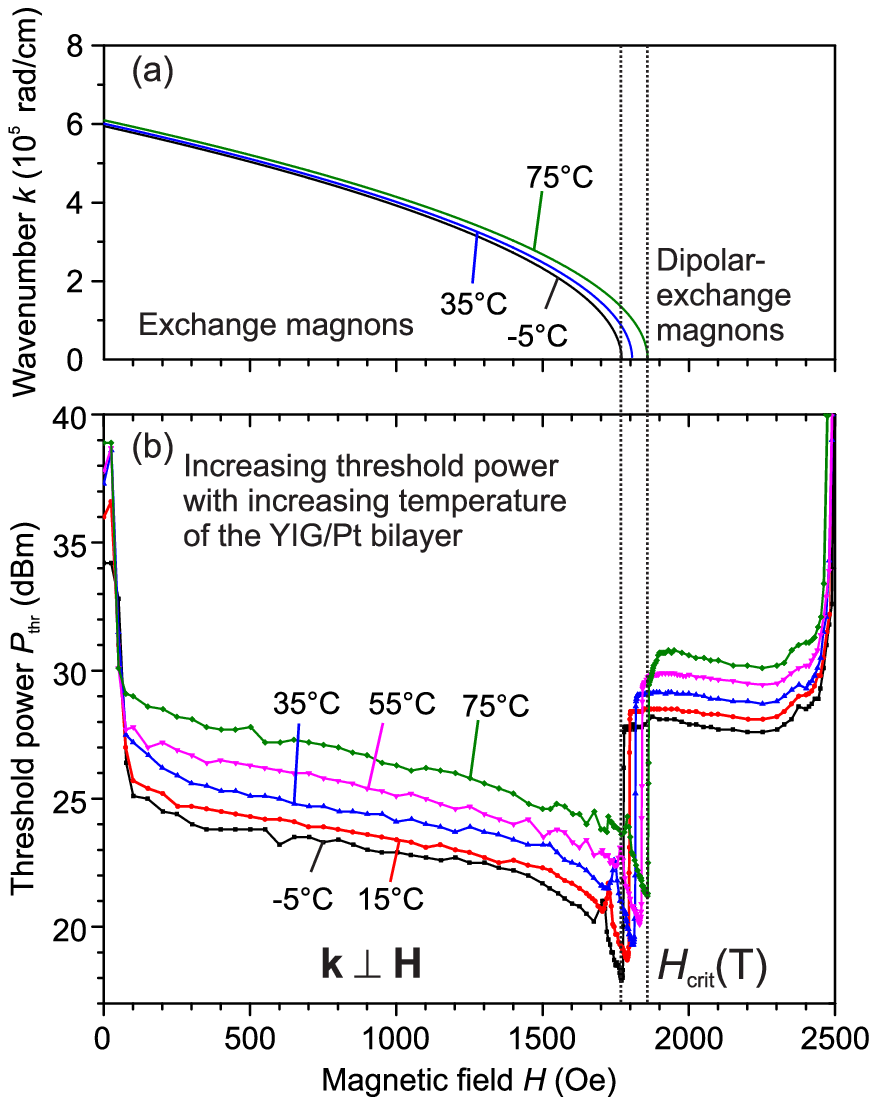}%
\caption{\label{bfly_const}(a) Excited wavenumbers for different temperatures with respect to the externally applied magnetic field. The spectrum was determined using the theory from Gurevich {\it et al.} \cite{Gurevich}. (b) Measured dependencies of the threshold power on the externally applied magnetic field for constant temperatures of the sample. Below the temperature dependent critical fields $H_{\mathrm{crit}}$ the generated magnon wavevector is oriented perpendicular to the applied magnetic field.}%
\end{figure}

In our experiments the threshold power is measured in a wide range of bias magnetic field values and thus is determined for a wide range of magnon wavenumbers (see Fig.~\ref{bfly_const}a). Firstly the temperature of the YIG/Pt bilayer was changed homogeneously in 20\,$^{\circ}$C steps from -5\,${^\circ}$C to 75\,${^\circ}$C. The results are shown in Fig.~\ref{bfly_const}b. In each case there is the typical dependence of the threshold power on the magnetic field \cite{Neumann2009}. Close to the ferromagnetic resonance (FMR) with wavenumber $k\to 0$, at the critical field $H_{\mathrm{crit}}$, the parametric excitation is most efficient due to the highest ellipticity in the precession. This ellipticity is caused by the dynamic stray field. Thus, the threshold pumping field strength is minimal. With decreasing external magnetic field in the range below the critical field ($H<H_{\mathrm{crit}}$), the threshold power increases slowly due to an increase in wavenumbers and a related decrease in the ellipticity of the excited spin waves \cite{Serga2012}. For magnetic fields little below $H_{\mathrm {crit}}$ a sharp peak in the threshold power can be found. It appears because of an increase of the threshold power due to interactions of the magnons with a transversal phonon mode. With increasing external magnetic field above the critical field the threshold power increases sharply \cite{Neumann2009}. The coupling of microwave photons to the corresponding magnons is reduced since the angle of the amplified spin waves to the parallel pump field is smaller than 90$^{\circ}$. Moreover, since these spin waves possess a nonzero group velocity component parallel to the external field, they flow out of the strongly localized amplification area of 50\,$\mu$m width on the pumping microstrip and therefore increase the effective damping. It is important to notice that there is only one distinct group of magnons parametrically excited for each magnetic field value. The spectral density of parametrically excited spin waves is typically of the order of several kilohertz \cite{Krutsenko}, what is much smaller than the typical FMR linewidth of a YIG film.

\begin{figure}[t]
\includegraphics[width=0.90\linewidth]{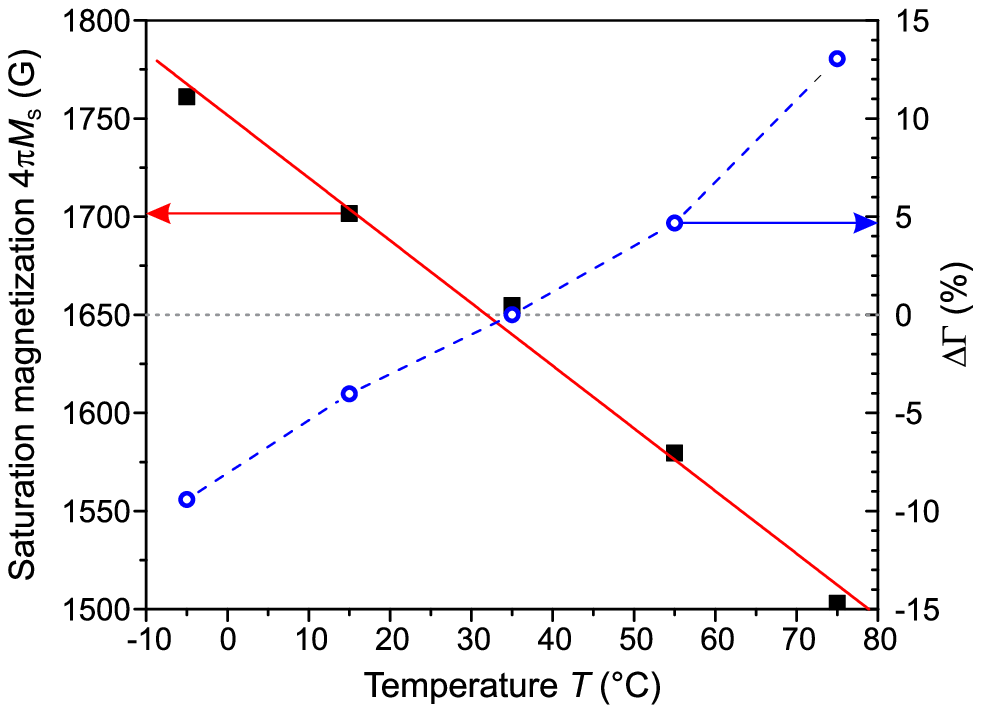}%
\caption{\label{satmag}Full squares: Calculated dependence of the saturation magnetization on the temperature of the YIG layer. The line is the result of a linear fit. Open circles: Temperature dependent change in relaxation parameter $\Delta\Gamma$ for the critical field relative to the corresponding relaxation parameter for the reference temperature of $T_{\mathrm{ref}}=$\,35\,$^\circ$C.}%
\end{figure}

Comparing the curves for different temperatures, we observe a change in the magnetic field position of the minimum. This occurs due to a change in the saturation magnetization by changing the temperature. The saturation magnetization is recalculated using Kittel's formula \cite{Kittel} with the values of the critical fields. The results are shown in Fig.~\ref{satmag}. The linear fit has a slope of $\Delta (4\pi M_{\mathrm S}) / \Delta T =$\,-3.2\,G$\cdot$K$^{-1}$, what is in good agreement to previous results \cite{Obry2012, Vogel2015, Algra1982}. Beside the shift of the critical field values in the threshold curve towards higher magnetic fields, a monotonous increase in the full range of the magnetic field in the threshold power is obvious. For each 20\,$^\circ$C temperature step we observe a difference of approximately 1\,dB in applied microwave power to reach the threshold value. According to Cherepanov {\it et al.} \cite{Cherepanov} the intrinsic damping in YIG strongly depends on the absolute temperature due to Kasuya-LeCraw processes\cite{Kasuya, Gurevich2}. A higher damping leads to a higher threshold power since it has to be overcome by a higher pumping field.

From the values of Fig.~\ref{satmag} we can estimate the change in the spin-wave relaxation parameter $\Gamma$ for the applied constant temperatures compared to the reference value at $T_{\mathrm{ref}}=$\,35\,$^{\circ}$C for the wavenumber $k\to 0$. We can determine the relative change in the relaxation parameters \cite{Gurevich} using

\begin{equation}
\Delta\Gamma = \frac{\Gamma(T)-\Gamma(T_{\mathrm{ref}})}{\Gamma(T_{\mathrm{ref}})}\cdot 100\%.
\end{equation}

Taking into account the connection between the relaxation parameter, the magnetization dependent coupling coefficient and the threshold field we can rewrite:

\begin{equation}
\Delta\Gamma = (\frac{M_{\mathrm S}(T)}{M_{\mathrm S}(T_{\mathrm{ref}})}\cdot \sqrt{\frac{P_{\mathrm {thr}}(T)}{P_{\mathrm {thr}}(T_{\mathrm{ref}})}}-1)\cdot 100\%
\end{equation}
with the corresponding threshold microwave power $P_{\mathrm{thr}}$. The resulting relative changes are shown in Fig.~\ref{satmag} as open circles. In the case of a temperature of 75\,$^{\circ}$C the relaxation frequency is 13.0\,\% higher, for -5\,$^{\circ}$C it is 9.4\,\% lower than the relaxation frequency at the reference temperature of 35\,$^{\circ}$C. This means that the change in damping by changing the absolute temperature is very pronounced. At the same time we can neglect the influence of the GGG substrate on the threshold power around temperatures of around 300\,K \cite{Danilov}. Therefore, all investigations on the pure influence of a temperature gradient must necessarily avoid any changes in the mean temperature of the system.

\begin{figure}
\includegraphics[width=0.90\linewidth]{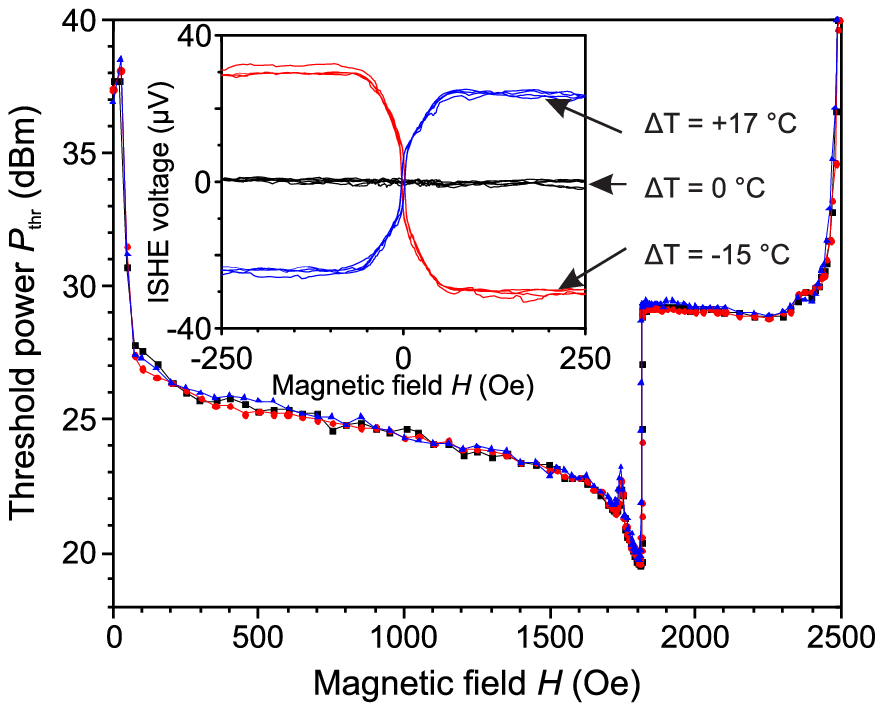}%
\caption{\label{bfly_grad}Measured dependencies of the threshold power on the externally applied magnetic field for different temperature gradients perpendicular to the sample plane. The mean temperature of T=35\,$^{\circ}$C of the sample is the same for all three curves. The inset shows the spin Seebeck voltage depending on the external magnetic field for the same temperature gradients.}%
\end{figure}

In the next step, temperature gradients were created across the sample thickness. In order to investigate the pure influence of a temperature gradient on the parametric pumping process it is important to keep the average temperature of the YIG layer, where magnons are excited, constant. The bottom Peltier element has been adapted so that we obtain the same mean temperature of the YIG film of 35\,$^{\circ}$C in every case. In first approach this mean temperature has been tuned by the electrical resistance of the Pt layer used as a thermometer and by an additional precise alignment of the FMR threshold minimum to the same magnetic field value in every case. The temperature on the top of the sample is either 17\,$^{\circ}$C higher or 15\,$^{\circ}$C lower than this base temperature, depending on the direction of the gradient. The dependence of the threshold power on the magnetic field for a homogeneous temperature of 35\,$^{\circ}$C is also used here for comparison.\\
A temperature gradient across the YIG/Pt bilayer interface leads to the longitudinal spin Seebeck effect (LSSE). In our system its existence has been proven by direct measurements of the LSSE voltage between the lateral edges of the Pt layer, see the inset in Fig~\ref{bfly_grad}. For opposite directions of the temperature gradient the measured voltages have opposite signs with a sign change at zero field\cite{SchreierISHE}.
For a homogeneous temperature of the sample of 35\,$^{\circ}$C no voltage is detected. At the same time the measurement of the threshold powers in the parametric pumping process reveal that there is no measurable influence of the longitudinal spin Seebeck effect on the spin-wave damping parameter (see Fig.~\ref{bfly_grad}). All three curves shown are on the same level within an experimental uncertainty.

In contrast to previous reports, where a broader range of magnons are excited, the parametric pumping process generates magnons of only one distinct magnon group within a very narrow bandwidth. Measurements of the ferromagnetic resonance in thick macroscopic YIG samples might show a temperature dependent heterogeneous broadening of the linewidth due to a relative shift of the many mode excitation in this process. An influence of the spin Seebeck effect on the well-defined mode excited by parametric pumping in our experiments was not found.\\
Recent theoretical investigations by Bender {\it et al.} \cite{Bender} show that the temperature difference at the interface between the electrons in the platinum layer and the magnons in the YIG layer needed to compensate the magnetic damping is at least in the order of the magnon energy $\hbar\Omega$. Hereby $\Omega$ is the magnon frequency, that is strictly limited to $2\pi\times$ 7\,GHz in our experiment. Therefore a visible change in the damping in the order of 5\% can be reached by a temperature difference $\Delta T_{\mathrm{mp}}$ of 16.8\,mK between the magnon temperature in YIG and the electron temperature in Pt. Our calculations using the theory of Xiao {\it et al.} \cite{Xiao2010, Schreier2013} applied for a mean temperature of 35\,$^{\circ}$C and the parameters of our experiment reveal a value of $\Delta T_{\mathrm{mp}}$ = 1.9\,mK. Thus, the influence of the spin Seebeck effect on the damping of parametric spin waves is even in theory too weak to be determined.
Lauer {\it et al.} \cite{Lauer2016} have found a spin-transfer torque by a spin polarized current created by the spin Hall effect that affects the parametric pumping threshold in thin YIG/Pt bilayers. Nevertheless an influence of the spin Seebeck effect could also not be observed in that report, what supports our findings.

In summary, the threshold power levels for the parallel parametric pumping process in YIG/Pt bilayers in a wide wavevector range have been investigated for different thermal configurations. It has been shown, that the threshold power strongly depends on the mean temperature of the YIG layer, whereas a temperature gradient does not change the threshold power as long as there is no change in the mean temperature of the YIG film with thickness in the order of micrometers. These tendencies are visible throughout the entire measured range of wavevectors. It has been confirmed that the longitudinal spin Seebeck effect is present, but finally its possible influence on the spin-wave damping is too weak to be determined in our experiments.


%
%

%

\begin{acknowledgments}
The authors thank Tobias Fischer as well as Bert L\"agel from the Nano\-struc\-turing Center (NSC) of TU Kaiserslautern for the preparation of the samples. Financial support by Deutsche Forschungsgemeinschaft (DFG) within priority program SPP1538 "Spin Caloric Transport" (projects VA 735/1-2 and SE 1771/4-2) is gratefully acknowledged.
\end{acknowledgments}


\end{document}